\documentstyle[12pt,epsfig,axodraw,a4]{article} 
\newcommand{\vs}{\vspace{-0.25cm}}
\begin{document} 

\begin{center}
{\Large{\bf Chiral SU(3) dynamics and \\ $\Lambda$-hyperons in the nuclear 
medium}}\footnote{Work supported in part by BMBF and GSI.}  
\bigskip

N. Kaiser and W. Weise\\
\medskip
{\small Physik-Department, Technische Universit\"{a}t M\"{u}nchen,
    D-85747 Garching, Germany}
\end{center}

\begin{abstract}
We present a novel approach to the density dependent mean field and the 
spin-orbit interaction of a $\Lambda$-hyperon in a nuclear many-body system, 
based on flavor-SU(3) in-medium chiral perturbation theory. The leading 
long-range $\Lambda N$-interaction arises from kaon exchange and from two-pion 
exchange with a $\Sigma$-hyperon in the intermediate state. The empirical 
$\Lambda$-nucleus potential depth of about $-28\,$MeV is well reproduced with 
a single cutoff scale,  $\bar \Lambda = 0.7\,$GeV, effectively representing
all short-distance (high-momentum) dynamics not resolved at scales
characteristic of the nuclear Fermi momentum. This value of $\bar\Lambda$ is
remarkably consistent with the one required to reproduce the empirical
saturation point of isospin-symmetric nuclear matter in the same
framework. The smallness of the $\Lambda$-nuclear spin-orbit interaction finds
a natural (yet novel) explanation in terms of an almost complete cancellation
between short-range contributions (properly rescaled from the known nucleonic 
spin-orbit coupling strength) and long-range terms generated by iterated 
one-pion exchange with intermediate $\Sigma$-hyperons. The small
$\Sigma\Lambda$-mass difference figures prominently in this context.
\end{abstract}
\medskip
PACS: 13.75.Ev, 21.65.+f, 21.80.+a, 24.10.Cn\\

\section{Introduction and framework}
The physics of $\Lambda$-hypernuclei has a long and well documented history 
\cite{1,2,bando}. It has at the same time raised questions of fundamental 
interest. The empirical single-particle energies of a $\Lambda$-hyperon bound
in hypernuclei are well described in terms of an attractive mean field about
half as strong as the one for nucleons in nuclei \cite{2}. In contrast, the 
extraordinary weakness of the $\Lambda$-nucleus spin-orbit interaction, as 
compared to the one in ordinary nuclei, has always been a puzzle. For example, 
recent precision measurements \cite{4} of E1-transitions from $p$- to 
$s$-shell orbitals of  a $\Lambda$-hyperon in $_\Lambda^{13} C$  give
a $p_{3/2} - p_{1/2}$ spin-orbit splitting  of only $(152 \pm 65)$\,keV, to be
compared with about 6\,MeV in ordinary p-shell nuclei. Even admitting a large
uncertainty in the hypernuclear case, the $\Lambda$-nuclear spin-orbit
interactions appears to be systematically weaker by at least an order of
magnitude than the N-nucleus spin-orbit force. Various attempts have been made
to understand this phenomenon \cite{5,6,pirner}. So far, standard calculations
using one-boson exchange $\Lambda N$-potentials \cite{7,dover,hiyama,millener}
tend to overestimate the $\Lambda$-nucleus spin-orbit splitting significantly.
More recently, in-medium effective field theory approaches have opened new 
perspectives for dealing with these issues in ordinary nuclei, both symmetric 
and asymmetric in isospin. In this work we take steps towards including 
strangeness in such a framework. 

Our calculation is based on the leading order chiral meson-baryon Lagrangian
in flavor-SU(3):   
\begin{equation} {\cal L}_{\phi B} = {\rm tr}(\bar B(\gamma_\mu(i\partial^\mu 
B+ [\Gamma^\mu,B])-M_B )B)+ D\, {\rm tr}(\bar B \gamma_\mu\gamma_5\{u^\mu,
B\}) + F\, {\rm tr}(\bar B\gamma_\mu\gamma_5 [u^\mu,B]) \,, \end{equation}
where the traceless hermitian $3\times 3$ matrix $B$ of Dirac-spinors 
represents the octet baryon-fields ($N,\Lambda,\Sigma,\Xi$), with mass $M_B$. 
The chiral connection $\Gamma^\mu = i[\xi^\dagger, \partial^\mu\xi]/2$ and the 
axial-vector quantity $u^\mu = i\{\xi^\dagger, \partial^\mu\xi\}/2$ generate 
interaction terms with the Goldstone-bosons ($\pi,K,\bar K,\eta$) collected in 
the SU(3) matrix $\xi = \exp(i\phi/2f)$. The parameter $f$ is identified with 
the weak pion decay constant $f_\pi = 92.4\,$MeV and $D$ and $F$ denote the 
SU(3) axial-vector coupling constants of the baryons. We choose as their 
values $D= 0.84$ and $F=0.46$. This leads to a $KN\Lambda$-coupling constant 
of $g_{KN\Lambda} = (D+3F)(M_N+M_\Lambda)/(2\sqrt{3} f_\pi) =14.25$ and a 
$\pi\Lambda \Sigma $-coupling constant of $g_{\pi\Lambda\Sigma} = D(M_\Lambda 
+M_\Sigma)/(\sqrt{3}f_\pi) =12.12$ which are both consistent with the 
empirical values summarized in Tab.\,6.3 and 6.4 of ref.\cite{coupl}. 
Furthermore, the pion-nucleon coupling constant has the value $g_{\pi N}= g_A 
M_N/f_\pi=  13.2$ with $g_A =D+F=1.3$. Apart from these three pseudovector 
$KN\Lambda$-, $\pi\Lambda\Sigma$- and $\pi NN$-couplings, no further
interaction terms from ${\cal L}_{\phi B}$ come into play to the order we are 
working here.
\begin{figure}
\begin{center}
\includegraphics[scale=1.]{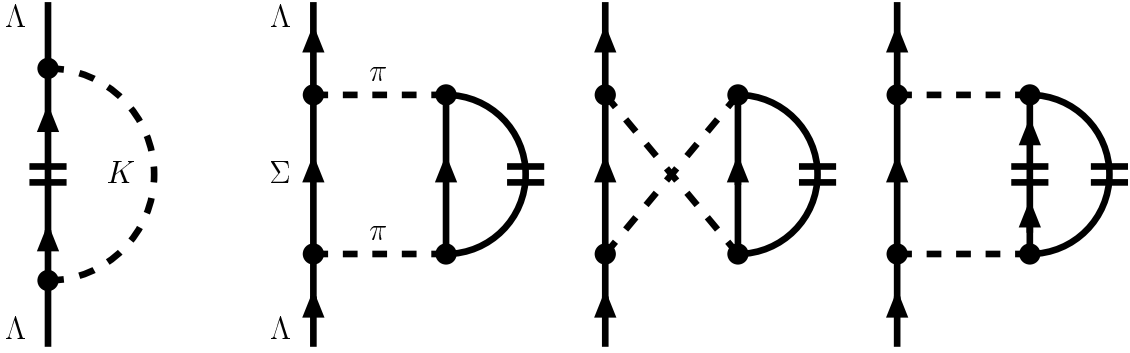}
\end{center}\vspace{-.2cm}
\caption{One-kaon exchange Fock diagram and two-pion exchange Hartree diagrams 
with $\Sigma$-hyperons in the intermediate state. The horizontal double-lines 
represent the filled Fermi sea of nucleons in the in-medium nucleon
propagator, $(\gamma\cdot p -M_N)[i(p^2 - M_N^2 + i\epsilon)^{-1} -
2\pi\delta(p^2 - M_N^2)\theta(p_0)\theta(k_f - |\vec{p}\,|)]$. The isospin
factors of the kaon- and pion-exchange diagrams are 2 and 6, respectively.}
\end{figure}
\bigskip
\section{$\Lambda$-nucleus single-particle potential}
Consider first the density dependent mean field $U_\Lambda(k_f)$ for a 
zero-momentum $\Lambda$-hyperon placed as a test particle in isospin-symmetric 
nuclear matter. The potential depth $U_\Lambda(k_{f0})$ at equilibrium density 
$\rho_0=0.16\,$fm$^{-3}$ determines primarily the spectra of medium heavy and 
heavy $\Lambda$-hypernuclei. We calculate the long-range contributions 
generated by the exchange of light Goldstone bosons between the 
$\Lambda$-hyperon and the nucleons in the filled Fermi sea. The only 
non-vanishing one-meson exchange contribution comes from the kaon-exchange 
Fock diagram in Fig.\,1, from which we obtain the following repulsive 
contribution to the $\Lambda$-nuclear mean-field:   
\begin{equation} U_\Lambda(k_f)^{(K)} = {(D+3F)^2 \over (6\pi f_\pi)^2} 
\Bigg\{k_f^3 -3 m_K^2  k_f+3 m_K^3 \arctan{k_f \over m_K} \Bigg\} +{\cal
O}(M_B^{-2}) \,, \end{equation}
with $m_K = 496\,$MeV the average kaon mass. At densities at and below
nuclear matter saturation density $\rho \leq 0.16\,$fm$^{-3}$ (corresponding
to Fermi momenta $k_f \leq 263\,$MeV) the one-kaon exchange can already 
be regarded as being of short range. The ratio $k_f/m_K \leq 0.53$ is small 
and the expression in curly brackets of eq.(2) is dominated by its leading 
term $3k_f^5/5m_K^2$ in the $k_f$-expansion.

Since one-pion exchange is excluded by isospin invariance, the leading
long-range interaction between the $\Lambda$-hyperon and the nucleons arises
from two-pion exchange. The corresponding two-loop diagrams with a
$\Sigma$-hyperon in the intermediate state are shown in Fig.\,1. The
small $\Sigma\Lambda$-mass splitting $M_\Sigma-M_\Lambda =77.5\,$MeV which 
comes into play here is comparable in magnitude to $k_{f0}^2/M_N$, twice the 
typical kinetic energies of the nucleons. Therefore it has to be counted 
accordingly in the energy denominator. Putting all pieces together we find 
from the second diagram in Fig.\,1 the following attractive contribution to
the $\Lambda$-nuclear mean field:     
\begin{equation} U_\Lambda(k_f)^{(2\pi)}= - {D^2 g_A^2 \over f_\pi^4}\!
\int\limits_{|\vec p_1| < k_f} \!\!\!{d^3 p_1 d^3 l\over (2\pi)^6} 
{ M_B\, {\vec l}\,^4 \over (m_\pi^2 +{\vec l}\,^2 )^2\, [ \Delta^2 
+{\vec l}\,^2 -\vec l \cdot \vec p_1] } +{\cal O}(M_B^{-1}) \,, \end{equation}
with $m_\pi = 138\,$MeV the average pion mass. The mean baryon mass $M_B =
(2M_N+M_\Lambda+M_\Sigma)/4 = 1047\,$MeV serves the purpose of averaging 
out differences in the kinetic energies of the various baryons involved. The 
relation $M_\Sigma-M_\Lambda = \Delta^2/M_B$ for the $\Sigma\Lambda$-mass 
splitting defines another small mass scale $\Delta$. Its magnitude $\Delta
\simeq 285\,$MeV is close to the Fermi-momentum $k_{f0}= 263\,$MeV at
saturation density. As it stands the $d^3l$-loop integral in eq.(3) is 
ultraviolet divergent. By subtracting $M_B/{\vec l}\,^2$ from the integrand it
becomes convergent and analytically solvable. After regularizing the 
remaining (structureless) linear divergence $\int_0^\infty dl\,1$ by a 
momentum cut-off $\bar \Lambda$ we get: 
\begin{equation} U_\Lambda(k_f)^{(2\pi)} = {D^2 g_A^2  M_B \over (2\pi f_\pi
)^4} \Bigg\{ - {4\bar \Lambda \over 3} \, k_f^3 + \pi m_\pi^3 k_f\,
\Phi\bigg({k_f^2\over m^2_\pi},{\Delta^2\over m_\pi^2} \bigg) \Bigg\} 
+{\cal O}(M_B^{-1})\,,  \end{equation}
with the function 
\begin{eqnarray}\Phi(u, \delta) &=& \delta-3 +{1 \over 4}(u-2\delta+6) 
\sqrt{4\delta-u} \nonumber \\ && + {2 \over \sqrt{u}}(2u+\delta^2-4\delta+3 )
\arctan{ \sqrt{u} \over 2+\sqrt{4\delta-u}} \,, \end{eqnarray} 
emerging from the combined loop and Fermi-sphere integration, where $u = k_f^2/
m_\pi^2$ and $\delta = \Delta^2/m_\pi^2$. The branch point
of the function $\Phi(u, \delta)$ at $k_f = 2\Delta$ is related to the 
kinematical threshold for the (on-shell) scattering process $\Lambda N\to 
\Sigma N$. This threshold is reached only at very high densities, $\rho \geq
1.4\,$fm$^{-3}$. Note that the decomposition in eq.(4) is optimal from
the point of view of separating effects from high and low mass scales. 
The (high-momentum) cut-off scale $\bar \Lambda$ effectively 
parameterizes the strength of an attractive $\Lambda N$-contact interaction. 
No dependence on the two low-mass scales, $m_\pi$ and $\Delta$, is left over 
in the corresponding term linear in the nucleon density $\rho= 2k_f^3/3\pi^2$. 
The third diagram in Fig.\,1 with crossed pion-lines corresponds to 
irreducible two-pion exchange between the $\Lambda$-hyperon and the nucleons. 
At leading order in the small momentum expansion it is exactly canceled by a 
$M_B$-independent contribution from the second diagram in Fig.\,1 (with 
parallel pion-lines). Note that the same exact cancellation between the planar
and crossed box graphs is at work in the isoscalar central channel of the
$2\pi$-exchange  NN-potential (for details see sect.\,4.2 in
ref.\cite{nnpap}). Thus we are  left with the Pauli-blocking correction to the
iterated pion-exchange diagram with intermediate $\Sigma$-states. From the last
diagram in Fig.\,1 we find the following repulsive contribution to the 
$\Lambda$-nuclear mean field:   
\begin{equation} U_\Lambda(k_f)^{(2\pi)}_{\rm med}=  {D^2 g_A^2 \over f_\pi^4}
\! \int\limits_{|\vec p_{1,2}| < k_f} \!\!\!{d^3 p_1 d^3 p_2 \over (2\pi)^6} 
 { M_B\, (\vec p_1-\vec p_2)^4 \over [m_\pi^2 +(\vec p_1-\vec p_2)^2]^2\,  
[\Delta^2 +{\vec p_2}^{\,2} -\vec p_1 \cdot \vec p_2] } +{\cal O}(M_B^{-1})\,.  
\end{equation}
After performing the angular integrations this expression reduces to an easily
manageable two-dimensional integral: 
\begin{eqnarray}U_\Lambda(k_f)^{(2\pi)}_{\rm med} &=& {D^2g_A^2M_Bm_\pi^4\over 
2(2\pi  f_\pi)^4} \int_0^u dx \int_0^u dy \,{1\over (2 \delta- 1-x+y)^2 } 
\, \Bigg\{ {4 (2\delta-1-x+y) \sqrt{x y} \over (1+x+y)^2-4x y} \nonumber \\ 
&&+ (2x-2y -4\delta+1)\ln {1+x+y + 2\sqrt{x y}\over 1+x+y -2 \sqrt{x y}}+ 
(2\delta-x +y)^2 \ln{\delta +y+\sqrt{x y}\over \delta +y-\sqrt{x y}}\Bigg\}\,,
\nonumber \\ \end{eqnarray} 
with the abbreviations $u = k_f^2/m_\pi^2$ and $\delta = \Delta^2/m_\pi^2$. 
\begin{figure}
\begin{center}
\includegraphics[scale=0.7]{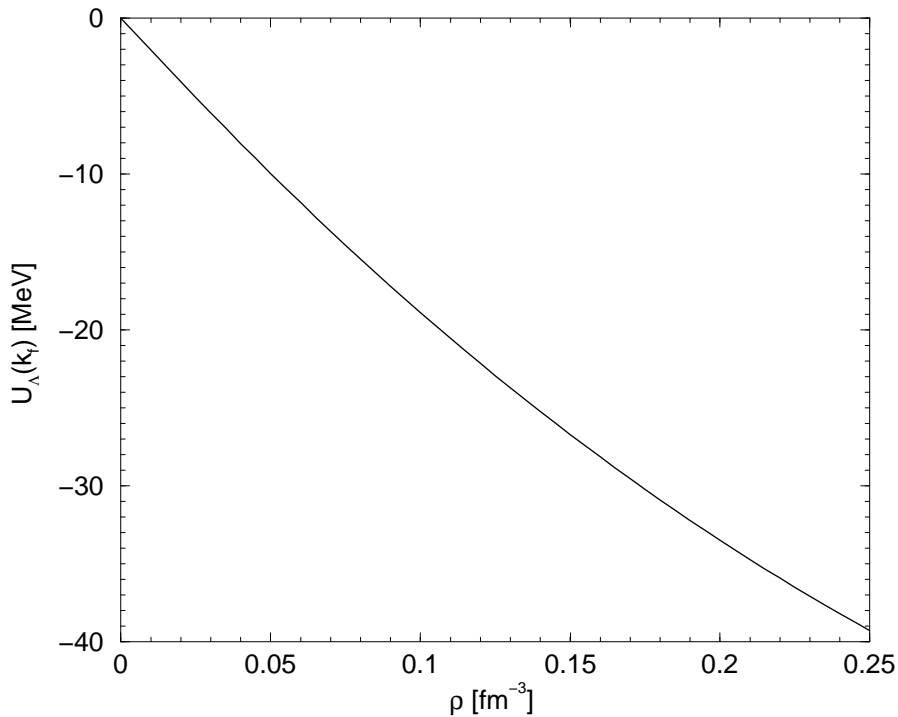}
\end{center}\vspace{-.8cm}
\caption{The mean field $U_\Lambda(k_f)$ of a $\Lambda$-hyperon in
isospin-symmetric nuclear matter versus the nucleon density  $\rho= 2k_f^3/3
\pi^2$. The cut-off scale  $\bar \Lambda$ has been adjusted to the value $\bar
\Lambda = 0.70\,$GeV.} 
\end{figure}

Summing up all terms, Fig.\,2 shows the calculated $\Lambda$-nuclear 
mean field $U_\Lambda(k_f)$ as a  function of the nucleon density $\rho=
2k_f^3/3\pi^2$. The cut-off scale has been adjusted to the value $\bar \Lambda 
= 0.7\,$GeV.  At normal nuclear matter density $\rho_0 = 0.16\,$fm$^{-3}$ 
(corresponding to $k_{f0} = 263\,$MeV) one finds $U_\Lambda(k_{f0})= (4.17 -
39.77  +7.46)\,$MeV$= -28.15\,$MeV, where the individual entries correspond to 
one-kaon exchange, iterated one-pion exchange with intermediate $\Sigma
N$-states, and the Pauli-blocking correction to the latter. Note that the
physically reasonable cut-off scale $\bar \Lambda = 0.70\,$GeV which
reproduces the empirical potential depth $U_\Lambda(k_{f0}) \simeq -28\,$MeV
\cite{1} is close to  $\bar \Lambda =0.65\,$GeV needed to reproduce the
empirical saturation point of isospin-symmetric  nuclear matter in the same
framework \cite{nucmat}. This is a remarkable and non-trivial feature. At the
present stage of our calculation all other possible contributions to the
$\Lambda$-nuclear mean field  $U_\Lambda(k_f)$ from $\pi K$-exchange, $\bar K
K$-exchange etc. are hidden in the adjusted value of the cut-off $\bar\Lambda
=0.70\,$GeV, or equivalently, in the contact term which encodes short-distance 
dynamics not resolved at momentum scales around $k_{f0}$. At the densities of 
interest all those effects can be regarded as being of short range nature and
therefore they are summarized by this single term linear in the nucleon 
density $\rho= 2k_f^3/3\pi^2$. 
\section{$\Lambda$-nucleus spin-orbit interaction}  
The empirical finding that the $\Lambda$-nucleus spin-orbit coupling is 
negligibly small in comparison to the strong spin-orbit interaction of 
nucleons in ordinary nuclei presents an outstanding problem in low-energy 
hadron physics. In relativistic scalar-vector mean field models a strong
tensor  coupling of the $\omega$-meson to the $\Lambda$-hyperon, equal and
of opposite sign to the vector-coupling, has been proposed as a possible 
solution \cite{6}. We shall demonstrate here that there is a more natural
source of cancellation in the hypernuclear many-body problem. 

The pertinent quantity to extract the $\Lambda$-nuclear spin-orbit coupling
is the spin-dependent part of the selfenergy of a $\Lambda$-hyperon 
interacting with  weakly inhomogeneous isospin-symmetric nuclear matter. Let
the $\Lambda$-hyperon scatter from initial momentum $\vec p-\vec q/2$ to final
momentum $\vec p+\vec q/2$. The spin-orbit part of the selfenergy in the 
weakly inhomogeneous medium is then \cite{uls}:
\begin{equation}\Sigma_{\rm spin} = {i \over 2} \,\vec \sigma \cdot (\vec q
\times \vec p\,) \, U_{\Lambda ls}(k_f)\,,
\end{equation}
where the density-dependent spin-orbit coupling strength $U_{\Lambda ls}(k_f)$
is taken in the limit of homogeneous nuclear matter (characterized by its
Fermi momentum $k_f$) and zero external $\Lambda$-momenta: $\vec p =\vec q
=0$. The more familiar spin-orbit Hamiltonian of the shell model follows from
eq.(8) by multiplication with a density form factor and Fourier 
transformation: 
\begin{equation} {\cal H}_{\Lambda ls} = U_{\Lambda ls}(k_{f0}) \,\,{1 \over
2r} {df(r)\over dr}\,\,  \vec \sigma \cdot \vec L\,.
\end{equation} 
Here $f(r)$ is the normalized nuclear density profile with $f(0)=1$, and $\vec
L = \vec r \times \vec p$ is the orbital angular momentum. For reference and 
orientation, consider first the frequently used simple model of isoscalar 
vector boson ($\omega$-meson) exchange between the $\Lambda$-hyperon and the
nucleon. The non-relativistic expansion of the vector (and tensor) coupling 
vertex between Dirac spinors of the $\Lambda$-hyperon gives rise to a 
spin-orbit term proportional to $i\,\vec \sigma \cdot (\vec q \times \vec p\,)
/4M_\Lambda^2$. Next one takes the limit of homogeneous nuclear matter (i.e. 
$\vec q=0$), performs the  remaining integral over the nuclear Fermi sphere
and arrives at the familiar result:  
\begin{equation}  U_{\Lambda ls}(k_f)^{(\omega)} = { G_V\over 2M^2_\Lambda} 
\, \rho \,, \end{equation}
linear in the nucleon density $\rho$. Here, $G_V=g_{\omega\Lambda}(1+2 \kappa_{
\omega \Lambda})g_{\omega N }/ m_\omega^2$ is a coupling strength of dimension 
(length)$^2$ which includes the $\omega$-baryon coupling constants, a possible 
tensor coupling of the $\omega$-meson to the $\Lambda$-hyperon with $\kappa_{
\omega\Lambda}$ the tensor-to-vector coupling ratio, and the $\omega$-meson
mass $m_\omega$. In the absence of  $\kappa_{\omega \Lambda}$ and assuming
that the $\omega$-meson does not couple to the strange quark in the
$\Lambda$-hyperon, one would naively expect $G_V$ to be $2/3$ of the 
corresponding piece of the NN-interaction for which $G_V \simeq 12$ fm$^2$. 
Evidently, an almost vanishing $\Lambda$-nuclear spin-orbit force would be 
difficult to understand at this stage unless one postulates a much reduced
coupling $g_{\omega\Lambda}$ or a sizeable and negative tensor-to-vector
coupling ratio $\kappa_{\omega\Lambda}$ \cite{6}. 

The important observation is now that iterated one-pion exchange with an
intermediate $\Sigma$-hyperon also generates a sizeable spin-orbit coupling, 
with a  sign opposite to that expected from the short-range interactions. The 
prefactor $i\,\vec \sigma \times \vec q$ is immediately identified by 
rewriting the product of $\pi \Lambda \Sigma$-interaction vertices
$\vec\sigma\cdot(\vec l- \vec q/2)\,\vec \sigma \cdot (\vec l + \vec q/2)$ at
the open baryon line in the $2\pi$-exchange process shown by the second
diagram in Fig.\,1. For all remaining parts of the 
iterated pion-exchange diagram one can then take the limit of homogeneous 
nuclear matter (i.e. $\vec q=0$). The other essential factor $\vec p$ emerges 
from the energy denominator $\Delta^2 +\vec l\cdot(\vec l-\vec p_1+\vec p\,)$. 
Keeping only the term linear in the external momentum $\vec p$ one finds from 
the second diagram in Fig.\,1 the following contribution to the
$\Lambda$-nuclear spin-orbit coupling strength: 
\begin{equation} U_{\Lambda ls}(k_f)^{(2\pi)}= - {2D^2 g_A^2 \over 3f_\pi^4}\!
\int\limits_{|\vec p_1| < k_f} \!\!\!{d^3 p_1d^3 l\over (2\pi)^6} { M_B\, 
{\vec l}\,^4 \over (m_\pi^2 +{\vec l}\,^2 )^2\, [ \Delta^2 
+{\vec l}\,^2 -\vec l \cdot \vec p_1]^2 }  \,. \end{equation}
This loop integral is convergent as it stands. It can be solved together
with the Fermi sphere integral in closed form: 
\begin{equation} U_{\Lambda ls}(k_f)^{(2\pi)} = {D^2 g_A^2 M_B m_\pi k_f\over
 24\pi^3 f_\pi^4}\, \,\Omega\bigg( {k_f^2\over m^2_\pi},{\Delta^2\over
m^2_\pi}\bigg) \,, \end{equation}
with the function: 
\begin{eqnarray}\Omega(u, \delta) &=& {1 \over u+(\delta-1)^2}\Big[ 6\delta-
2\delta^2 -4 -3u +(\delta^2-3\delta+2 +u )\sqrt{4\delta-u}\Big]  \nonumber \\ 
&& + {4\over \sqrt{u}}(2-\delta) \arctan{\sqrt{u}\over 2+\sqrt{4\delta-u}} 
\,. \end{eqnarray}
The Pauli-blocking correction to the $\Lambda$-nuclear spin-orbit coupling
strength generated by iterated pion-exchange is calculated in the same way:  
\begin{equation} U_{\Lambda ls}(k_f)^{(2\pi)}_{\rm med}=  {2 D^2 g_A^2 \over 
3 f_\pi^4} \! \int\limits_{|\vec p_{1,2}| < k_f} \!\!\!{d^3 p_1 d^3 p_2 \over 
(2\pi)^6} { M_B\, (\vec p_1-\vec p_2)^4 \over [m_\pi^2 +(\vec p_1-\vec p_2)^2
]^2\, [\Delta^2 +{\vec p_2}^{\,2} -\vec p_1 \cdot \vec p_2]^2 }\,,
\end{equation} 
and after performing the angular integrations it turns into the numerically
easily manageable form:
\begin{eqnarray}U_{\Lambda ls}(k_f)^{(2\pi)}_{\rm med} &=&{D^2g_A^2M_B m_\pi^2
\over 12(\pi  f_\pi)^4} \int_0^u dx \int_0^u dy \,{1 \over (2\delta-1-x+y)^2 } 
\,  \Bigg\{ {2 \sqrt{xy} \over (1+x+y)^2-4x y}\nonumber \\ &&+{(2\delta-x+y)^2
  \sqrt{xy}\over 2(\delta+y)^2-2x y}  +{2\delta -x+y \over 2\delta-1-x+y} 
\ln{(\delta + y +\sqrt{xy})(1+x+y - 2\sqrt{xy})\over (\delta + y -\sqrt{xy}) 
(1+x+y + 2\sqrt{xy})}\Bigg\} \,. \nonumber \\ \end{eqnarray}
The two terms, eqs.(12,15), are model independent in the sense that they do 
not require any regularization. Their input parameters (couplings constants 
and masses) are physical quantities and thus uniquely fixed. Note that these 
spin-orbit couplings are not relativistic effects: they are even proportional 
to the baryon mass  $M_B$. This large scale enhancement factor originates from 
the energy denominator of the iterated pion-exchange diagram. The expressions 
in eqs.(12,15) constitute the unique long-range $\Lambda$-nuclear spin-orbit
interaction. The summed contributions from these $2\pi$-exchange processes are
shown by the lower solid line in Fig.\,3. They are of comparable magnitude but
of opposite sign with respect to the short-range pieces mentioned earlier. The
short-range part of the $\Lambda$-nuclear spin-orbit interaction results from
a variety of processes, one of them being the isoscalar-vector exchange piece
discussed previously. We relate the short-distance spin-orbit coupling of the
$\Lambda$-hyperon to the corresponding one of the nucleon as follows:  
\begin{equation} U_{\Lambda ls}(k_f)^{(\rm sh)}=  C_{ls}{M_N^2 \over 
M_\Lambda^2} \,  U_{N ls}(k_f)^{(\rm sh)}\,. \end{equation} 
The factor $(M_N/M_\Lambda)^2$ results from the replacement of the nucleon by 
a  $\Lambda$-hyperon in these relativistic spin-orbit terms.  The coefficient 
$C_{ls}$ parameterizes the ratio of the relevant coupling strengths. The upper 
limit expected from naive quark model considerations is $C_{ls} = 2/3$. 
Estimates from a QCD sum rule analysis of the Lorentz scalar and vector mean 
fields of a $\Lambda$-hyperon in a nuclear medium \cite{jin} indicate that 
$C_{ls}$ is  smaller than its naive quark model value, partly as consequence 
of flavor-SU(3) breaking. For nucleons in nuclei, large Lorentz scalar and 
vector mean fields with their in-medium behavior governed by QCD sum rules can 
explain the strong spin-orbit coupling in calculations which combine these 
mean fields with the long- and intermediate range attraction  provided by 
perturbative chiral pion-nucleon dynamics \cite{finelli}. Consider now the
leading terms of the  vector selfenergies $\Sigma_V$, linear in the quark
density $\langle u^\dagger u\rangle =\langle d^\dagger d\rangle = 3\rho/2$, in
a nuclear medium with only u- and d-quarks. From ref.\cite{jin} one estimates
roughly $\Sigma_V(\Lambda)/\Sigma_V(N) \simeq 1/2$ for the ratio of vector 
mean fields experienced by a $\Lambda$-hyperon and a nucleon. Corrections from
in-medium condensates of higher dimension tend to reduce this ratio
further. For the Lorentz scalar mean fields, the QCD sum rule results are
subject to larger uncertainties due to the unknown contributions from
four-quark condensates. On the other hand, at least some of these
contributions are accounted for by our explicit treatment of two-pion exchange 
processes. We therefore assume as a reasonable estimate, guided by \cite{jin}, 
a factor $C_{ls} \simeq 0.4 - 0.5$ for both the scalar and vector selfenergies 
which act coherently to produce the short-range spin-orbit force and cancel in 
the spin-averaged single-particle potential $U_\Lambda(k_f)$. In practice, we
shall vary $C_{ls}$ between 1/2 and 2/3. 
\begin{figure}
\begin{center}
\includegraphics[scale=0.7]{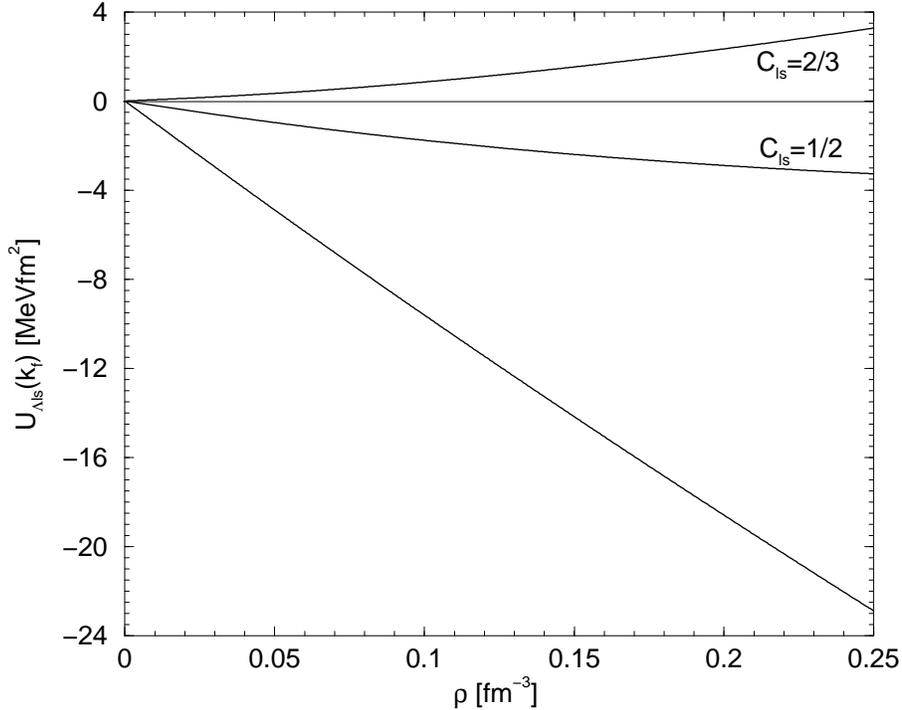}
\end{center}\vspace{-.8cm}
\caption{The spin-orbit coupling strength $U_{\Lambda ls}(k_f)$ of a 
$\Lambda$-hyperon in isospin-symmetric nuclear matter versus the nucleon 
density $\rho= 2k_f^3/3\pi^2$. The lower curve shows the long-range 
contribution from iterated $1\pi$-exchange with $\Sigma$-hyperons in the 
intermediate state. The two upper curves include in addition the short-range 
contribution,  $U_{\Lambda ls}(k_f)^{(\rm sh)} = 24.8\,C_{ls}
\,$MeVfm$^2\cdot \rho/\rho_0$, with $C_{ls} = 2/3$ and $1/2$.} 
\end{figure}

For the further discussion we take the value $U_{N ls}(k_{f0})^{(\rm sh)} = 
35\,$MeVfm$^2$ of the nucleonic spin-orbit coupling strength from shell model
calculations \cite{bohr}. Phenomenological Skyrme forces 
\cite{sly} give approximately the same value $U_{N ls}(k_{f0})^{(\rm sh)} = 
3\rho_0 W_0/2 \simeq 30\,$MeVfm$^2$ (with $W_0 = 124\,$MeVfm$^5$ the
spin-orbit parameter in the Skyrme phenomenology). The lower curve in Fig.\,3 
shows the $\Lambda$-nuclear spin-orbit coupling strength generated by iterated
pion exchange with $\Sigma$-hyperons in the intermediate state, as a function
of the nucleon density $\rho = 2k_f^3/3\pi^2$. The upper curves include in
addition the short-range contribution  $U_{\Lambda ls}(k_f)^{(\rm sh)} =
24.8\,C_{ls}\,$MeVfm$^2 \cdot \rho/\rho_0$ which is obtained via eq.(16) from
the empirical nucleonic spin-orbit coupling strength with $C_{ls}$ taken at
the values 2/3 and 1/2. At nuclear matter saturation density $\rho_0= 
0.16\,$fm$^{-3}$ one finds $U_{\Lambda ls}(k_{f0}) = (24.8\,C_{ls} - 16.70 
+1.64)\,$MeVfm$^2$, where the individual entries correspond to the 
short-range term, the contribution from iterated $1\pi$-exchange and the 
Pauli-blocking correction to the latter. One observes a strong cancellation 
between the short- and long-range contributions. This so far unnoticed balance 
between sizeable ''correct-sign'' and ''wrong-sign'' spin-orbit terms for the 
$\Lambda$-hyperon offers a novel and natural explanation for the empirically 
observed small spin-orbit splittings in $\Lambda$-hypernuclei, although with 
still persisting uncertainties in the short-range contribution.

It is important to note that such a ''wrong-sign'' spin-orbit interaction from
iterated one-pion exchange (entirely through the second-order tensor force) 
exists indeed also for nucleons (see Fig.\,4 in ref.\cite{efun}).\footnote{One
should note, however, that the one-pion exchange tensor force is too strong at
intermediate and short distances, so that its effect on the spin-orbit 
coupling, as calculated in this work, represents an upper limit in magnitude.} 
It has however been found recently that three-body spin-orbit forces generated
by $2\pi$-exchange with virtual $\Delta(1232)$-isobar excitation compensate 
this contribution to a large extent \cite{delta3so} (see Fig.\,2 therein), 
leaving room for additional short-distance contributions. In the case of the 
$\Lambda$-hyperon the analogous three-body effects with virtual $\Delta(1232)
$-isobar excitation are not possible and therefore the sizeable ''wrong-sign'' 
spin-orbit interaction generated by iterated pion-exchange becomes visible. 
The absence of analogous three-body mechanisms for the $\Lambda$-hyperon
becomes immediately clear by inspection of the relevant $2\pi$-exchange 
diagram in Fig.\,4. Replacing the external nucleon by a
$\Lambda$-hyperon introduces as the intermediate state on the open baryon line
a $\Sigma$-hyperon for which there exists no filled Fermi sea. The so far
emerging picture of the nuclear spin-orbit interaction is a rather intriguing 
one. The spin-orbit coupling of nucleons in nuclei is predominantly of 
short-range origin because the long-range $2\pi$-exchange components find a 
mechanism of self-cancellation. The smallness of the $\Lambda$-nuclear 
spin-orbit coupling, on the other hand, reveals the existence of a long-range 
$2\pi$-exchange component of the ''wrong sign''. 
\begin{figure}
\begin{center}
\includegraphics[scale=1.1]{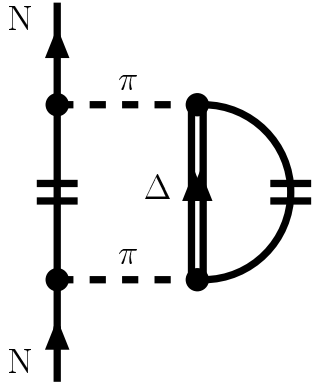}
\end{center}\vspace{-.4cm}
\caption{Three-body diagram of two-pion exchange with virtual 
$\Delta(1232)$-isobar excitation. For a nucleon it generates a sizeable 
three-body spin-orbit force of the ''right sign''. The horizontal double-lines 
symbolize the filled Fermi sea of nucleons. The analogous diagram does not
exist for a $\Lambda$-hyperon.}
\end{figure}

\section{Concluding remarks}
In summary we have calculated the density dependent
$\Lambda$-nuclear mean-field $U_\Lambda(k_f)$ in the framework of 
SU(3) chiral perturbation theory. The leading order contributions emerge from 
kaon-exchange and iterated pion-exchange with $\Sigma$-hyperons in the 
intermediate state. The empirical potential depth $U_\Lambda(k_{f0}) \simeq 
-28\,$MeV is well reproduced with a cut-off scale $\bar \Lambda = 
0.70\,$GeV, equivalent to a contact interaction, which represents effectively 
all short-range dynamics not "resolved" at scales characteristic of the 
nuclear Fermi momentum. The anomalously small $\Lambda$-nuclear spin-orbit 
interaction finds a novel and natural explanation in terms of the strong 
cancellation between short-range contributions (roughly estimated from the 
empirical nucleonic spin-orbit coupling strength, admittedly with large 
uncertainties) and long-range contributions generated by iterated
pion-exchange with $\Sigma$-hyperons in the intermediate state. The 
exceptionally small $\Sigma\Lambda$-mass splitting $M_\Sigma - M_\Lambda = 
77.5\,$MeV prominently influences the long-range iterated $1\pi$-exchange 
effects.

\bigskip

Acknowledgements: We thank R. Furnstahl, A. Gal and G. Lalazissis for
informative  discussions.

\end{document}